\documentclass{article} %%% use \documentstyle for old LaTeX compilers

\usepackage[english]{babel} %%% 'french', 'german', 'spanish', 'danish', etc.
\usepackage{amssymb}
\usepackage{amsmath}
\usepackage{amsthm}
\usepackage{txfonts}
\usepackage{mathdots}
\usepackage[pdftex]{graphicx} %%% use 'pdftex' instead of 'dvips' for PDF output
\begin{document}

%\selectlanguage{english} %%% remove comment delimiter ('%') and select language if required

\noindent 

\noindent 

\noindent 

\noindent 
\begin{center}
\noindent \textbf{\Large{A network theoretic study of ecological connectivity in Western Himalayas}}

\vspace{0.5 cm}

\noindent Shashankaditya Upadhyay${}^{1}$, Arijit Roy${}^{2}$, M. Ramprakash${}^{2}$, Jobin Idiculla${}^{3}$, A. Senthil Kumar${}^{2}$,
\\
\noindent Sudeepto Bhattacharya${}^{1,*}$
\\

\noindent ${}^{1}$ Department of Mathematics, School of Natural Sciences, Shiv Nadar University, Post Office Shiv Nadar University, Gautam Buddha Nagar, Greater Noida 201 314, Uttar Pradesh, India

\noindent ${}^{2}$ Indian Institute of Remote Sensing, Indian Space Research Organization, 4, Kalidas Road, Dehradun 248 001, Uttrakhand, India

\noindent ${}^{3}$ Department of Mathematics, School of Mathematics and Computer science, Central University of Tamil Nadu, Thiruvarur 610 101, Tamil Nadu, India

\noindent ${}^{*}$ Corresponding author. Email: sudeepto.bhattacharya@snu.edu.in
\end{center}
\vspace{1cm}
\noindent \textbf{Abstract}

\noindent Network theoretic approach has been used to model and study the flow of ecological information, growth and connectivity on landscape level of anemochory plant species \textit{Abied pindrow, Betula utilis} and \textit{Taxus wallichiana} in the Western Himalaya region. A network is formally defined and derived for seed dispersion model of aforementioned species where vertices represent habitat patches which are connected by an edge if the distance between the patches is less than a threshold distance. We define centrality of a network and computationally identify the habitat patches that are central to the process of seed dispersion to occur across the network. We find that the network of habitat patches is a scale-free network and at the same time it also displays small-world property characterized by high clustering and low average shortest path length. Due to high clustering, the spread of species is locally even as seed disperse mutually among the member vertices of a cluster. Also since every vertex is only a short number of steps away from every other vertex, the species rapidly covers all the habitat patches in the component. Also due to presence of hubs in the network the spread of species is greatly boosted whenever the species establish and thrive in a hub patch and disperse to adjacent patches. However, the network is not modular due to geographical constraints, and is negatively assortative as the high degree vertices are connected to vertices of low degree.

\vspace{0.5 cm}

\noindent \textbf{AMSC: 05C82, 90B10}

\vspace{0.5cm}

\noindent

\vspace{0.5 cm}

\noindent \textbf{\textsc{1. Introduction}}

\noindent 

\noindent Almost all the natural ecosystems on earth are experiencing degradation and destruction due to human activities. In many of the landscapes, large tracts of contiguous forests no longer exists and remnant natural habitats occur as a mosaic of large and small forest patches. For sustaining the ecosystem processes, a robust movement of energy, information and materials is a prerequisite. But in a fragmented landscape, the forest patch connectivity is important for unhindered movement of energy, information and materials. Understanding the functional connectivity of the patches, can provide invaluable information on the conservation policy to the followed as these provide the stepping stone for dispersal and movement of various species [1][2][3][4]. 

\noindent One of the most formal means to explore, examine and understand the essential structural and functional dynamics of ecological complexity is provided by a complex network-theoretic (and graph-theoretic) approach to ecosystem analysis [5][6][7][8][9]. Our work is essentially in the direction to realize the western Himalaya forest ecosystem as a complex system. For this we identify the individual habitat patches (entities) of the focal species and try to gain an understanding of the interaction they undergo, thus giving rise to various characteristics of the forest system that they are embedded in. Use of graph-theoretic network analysis provides relevant quantitative measures to analyse the landscape and responds directly to the level of isolation of the forest fragments in a changing landscape [10][11][12][13][14][15][16].

\noindent Landscape connectivity is primary in promoting the survival and vitality of species through flow of ecological information in the form of organism movement, seed dispersal and other ecological processes [17]. Maintaining connectivity and mitigating the fragmentation of habitat may be critical for landscape process such as gene flow and dispersal [18]. Advancements in ecological network flow modelling has led to the foundation of Ecological/Ecosystem Network Analysis (ENA), a method to holistically analyze environmental interactions [19][20][21][22][23][24][25][26][27][28]. We apply concepts from the theory of complex networks to study the level of connectivity of \textit{Abies pindrow, Betula utilis }and \textit{Taxus wallichiana} to find habitat patches which are critically important in conservation of connectivity pattern over last three decades (1985 -- 2014) in Western Himalayas. One of the ways to identify potential areas of spread of anemochory floral species is through modelling their distribution. Spatial modelling for species distribution is frequently being used for management of natural resources by environmentalists [29]. The importance of dispersal and movement of species through the landscape have been emphasized in the developments in metapopulation biology and landscape ecology, with species populations interacting dynamically through landscape-scale movements [30][31][32][33]. The restriction of gene flow and dispersal results in isolated populations, which leads to the loss of genetic material [34]. Graph theoretic approach to landscape analysis thus allows us to assess the importance of individual landscape elements and to guide conservation and restoration efforts [35][36].

\noindent Networks comprise generic representation of complex systems in which the underlying topology is a graph. Networks are effectively used to mathematically model empirical data from real world problems where the relationship between given components is of importance and may evolve with time.  Formally a network \textit{N} is a four tuple $ N = (V_{\lambda },\ E_{\lambda },\ {\psi }_{\lambda },\ \mathrm{\Lambda }$) along with an algorithm \textit{A} such that for$\ \mathrm{\Lambda }\ \neq \ \emptyset $, $i\in \mathrm{\Lambda },\ V_{\lambda }$ is a set of vertices $V_i\ ,\ E_{\lambda }$ is a set of edges $E_i\ ,\ {\psi }_{\lambda }$ is incidence function ${\psi }_{i\ }:E\to [V]^2$ where [V]${}^{2 }$is the set of not necessarily distinct unordered pairs of vertices such that $\left(V_i,\ E_i,\ {\psi }_i\right)\ $is a graph given by the algorithm \textit{A(i).} The incidence function $\psi $ provides structure to a graph by associating to each edge an unordered pair of vertices in the graph as $\psi \left(e\right)=\left\{v_i,\ v_j\right\}:v_i$, $v_j\in V,\ \forall \ e\in E\ \subseteq [V]^2$. Here $i$${}_{ }$ is the temporal component by virtue of which a network can evolve as per the given algorithm \textit{A}.

\noindent A network is thus an empirical object while the underlying graph that represents the network is an algebraic object such that an unlabeled graph represents an isomorphism class of otherwise labeled graphs. We call a network as static network if the temporal component $\mathrm{\Lambda }$ consist of a single element \textit{i}, otherwise the network is a dynamic network. For the purpose of our work in this paper, we define an ecological network as a network $N$ in which $V_{\lambda}$ is the set of habitat patches for the target species/population, and $E_{\lambda}$ is the set of flows of ecological matter between two distinct habitat patches.

\noindent The seed dispersion of \textit{Abies pindrow, Betula utilis }and \textit{Taxus wallichiana} forms a dynamic ecological network$\ N = (V_{\lambda },\ E_{\lambda },\ {\psi }_{\lambda },\ \mathrm{\Lambda }$) with $\mathrm{\Lambda }$ consisting of three periods over the years 1985, 1995 and 2005. Here the vertices of underlying graphs represent habitat patches of these species that are connected by edges whenever the distance between two habitat patches is less than three hundred meters. This is so because three hundred meters is considered the threshold up to which these wind dispersed (anemochory) species can disperse. 

\vspace{0.5 cm}

\noindent 

\noindent 

\noindent 

\noindent \textbf{\textsc{2. Study area}}

\noindent The study has been carried out in the Western Himalayan region of India constituting the states of Uttarakhand, Himachal Pradesh and Jammu and Kashmir. The study area lies between 28${}^\circ$43'N to 37${}^\circ$05'N latitude and 72${}^\circ$31'E to 81${}^\circ$03'E longitude. It has a total geographical area of 3,31,382 km${}^{2}$of which Himachal Pradesh, Jammu \& Kashmir and Uttarakhand covers 55,673 km${}^{2}$, 2,22,236 km${}^{2}$ and 53,483 km${}^{2}$ respectively. The altitude varies from foothills of Himalaya ca. 50 meters till 5,500 meters. The terrain is so diverse that it includes plains, undulating hills and high mountains[37]. The average annual rainfall is about 1,800 mm, 600 to 800 mm and 1,550 mm in Himachal Pradesh, Jammu \& Kashmir and Uttarakhand respectively. Temperature varies from sub-zero to 40${}^\circ$C. The region witnesses temperate climate except in the plains area where the climate is tropical. As per the State of Forest report (2011), the recorded forest area of Himachal Pradesh is 37,033 km${}^{2}$ which is 66.52\% of its geographical area. Reserve forests constitute 5.13\%, Protected forests 89.46\%, and unclassed forests 5.41\% of the recorded forest area. About two thirds of the state's recorded forest area is under permanent snow, cold deserts or glacier which is not conducive for the growth of trees. The recorded forest area of Jammu \& Kashmir is 20,230 km2 which is 9.1\% of the geographical area. Reserve forests constitute 87.21\%, protected forests 12.61\% and unclassed forests 0.81\% of the recorded forest area. The recorded forest area of Uttarakhand is 34.651 km2 which is 64.79\% of its geographical area. Reserve forests constitute 71.11\%, protected forests 28.52\%, and unclassed forest 0.35\% of the recorded forest area.

\vspace{0.5 cm}

\noindent 

\noindent \textbf{\textsc{3. Materials and methods}}

\noindent 

\noindent Land use and land cover (LULC) of the whole Western Himalayas generated for 1985, 1995 and 2005 that includes the three states of Himachal Pradesh, Jammu Kashmir and Uttarakhand was generated using multispectral satellite data (Landsat MSS, IRS LISS II and LISS III) at 1: 50,000 using onscreen visual interpretation. The maps were verified for accuracy using field data as well as high resolution Google Earth images for all the three time periods. The forest patches were subsequently extracted from the LULC. Since for a viable population of the target species a minimum patch size is required which is around 5 hactare, hence forest habitats with less than 5 hectare area were removed for the analysis. 

\noindent Using the vector files prepared of the LULC of three time periods, two ASCII files named vertex file and distance file were generated using a open source software Confer 2.5 for each of the years (1985, 1995 and 2005). These two ASCII files were the inputs for the network analysis as the vertex and distance file together gives an adjacency list for the three years, where vertices are forest patches and distance between them is given. And thus two vertices are considered adjacent in the network if the distance between them is less than the threshold distance of three hundred meters.
\\
\noindent 

\noindent \textbf{3.1 Computation of Network centrality}

\noindent 

\noindent Centrality indices are to quantify the intuitive notion that some vertices are relatively important and central than others in the given network. Let \textit{G} be a weighted or unweighted, directed or undirected multigraph and let \textit{X} represent the set of the set of vertices or edges of \textit{G} respectively. A real valued function \textit{s} is called a structural index if and only if following condition is satisfied
\begin{equation} \label{GrindEQ__1_} 
\forall x\in X,\ :G\ \cong H\ \Rightarrow \ s_G\left(x\right)=\ s_H\left(\phi \left(x\right)\right),\  
\end{equation} 
where $s_G\left(x\right)$ denotes the value of s(x) in G [38]. A real valued function is called a centrality index if and only if the following conditions are satisfied
\[\forall x,\ y,\ z\in X,\ f\left(x\right)\le f\left(y\right)or\ f\left(y\right)\le f\left(x\right)\] 
\[f\left(x\right)\le f\left(y\right)\ and\ f\left(y\right)\le f\left(z\right)\ then\ f\left(x\right)\le f(z)\] 
\begin{equation} \label{GrindEQ__2_} 
f\left(x\right)\le f\left(y\right)and\ f\left(y\right)\le f\left(x\right)\ then\ f\left(y\right)=f(x) 
\end{equation} 
that is \textit{f} induces a total order on X. By this order we can say that $x\in X$ is at least as central as $y\in X\ $with respect to a given centrality \textit{c} if $c\left(x\right)\ge c(y)$. Thus every centrality index is a structural index but every structural index need not be a centrality index. Several centrality indices have been proposed by researchers from various fields over time. Indices like degree centrality, betweenness centrality, closeness centrality, eigenvector centrality and subgraph centrality are discussed here as well as computed for our network.

\noindent 

\noindent The degree of a vertex \textit{u }in a graph \textit{G }is the number of edges incident on that vertex. The degree centrality of a vertex is given by
\begin{equation} \label{GrindEQ__3_} 
C_D\left(u\right)=d(u) 
\end{equation} 
where \textit{d}(\textit{u}) is the degree of the given vertex [39]. It can be computed as marginals of adjacency matrix \textit{A}. $C^i_D=\mathrm{\ }{\mathrm{\Sigma }}_j\ A_{ij}$. As per degree centrality a vertex is more central the more it is directly connected to other vertices. Thus degree centrality is understood as a measure of immediate influence in a network.

\noindent Betweenness centrality characterizes the importance of a vertex when information is passed through a given pair of vertices [40]. Betweenness centrality measures the number of times the information passes through a vertex \textit{k }by a shortest path between vertices \textit{i }and \textit{j}. Let ${\delta }_{st}\left(v\right)$ denotes the fraction of shortest paths between vertices \textit{s} and \textit{t} that contains \textit{v} such that ${\delta }_{st}\left(v\right)=\ \frac{{\sigma }_{st}(v)}{{\sigma }_{st}}$, where ${\sigma }_{st}\ $is the number of all shortest path between s and t. The betweenness centrality of a vertex is given by
\begin{equation} \label{GrindEQ__4_} 
C_B\left(v\right)={\mathrm{\Sigma }}_{s\neq t\neq v}{\delta }_{st}(v) 
\end{equation}

\noindent Closeness Centrality consider a vertex more central if the total sum of distance from the given vertex to all other vertices is minimum. The closeness centrality is given by
\begin{equation} \label{GrindEQ__5_} 
C_C\left(u\right)=\ \frac{1}{\sum_{v\in V}{d(u,v)}}.      
\end{equation} 
This measure is useful in the applications where a vertex is ranked higher if in terms of geodesic distance the vertex is close to most other vertices in the network.

\noindent 

\noindent Eigenvector centrality preserves the idea that a vertex is central if either it has many adjacent vertices or it is adjacent to vertices that have many adjacent vertices and hence are more central in the network [41]. Eigenvector centrality is given by the equation
\begin{equation} \label{GrindEQ__6_} 
Ax={\lambda }_1x 
\end{equation} 
where ${\lambda }_1$ is the largest eigenvalue of adjacency matrix \textit{A}.

\noindent 

\noindent Subgraph centrality of a vertex is the weighted sum of closed walks of different lengths in a network starting and ending at the given vertex [42]. Subgraph centrality is given by the equation
\begin{equation} \label{GrindEQ__7_} 
C_S\left(u\right)=\ \sum^{\infty }_{k=0}{\frac{{\mu }_k(i)}{k!}} 
\end{equation} 
where ${\mu }_k(i)$ is the closed walks of length k starting and ending on vertex \textit{i.}
\\
\noindent 

\noindent 

\noindent \textbf{3.2 Identification of scale-free networks}

\noindent 

\noindent A given quantity \textit{x }is said to obey a power law if its probability distribution is given by
\begin{equation} \label{GrindEQ__8_} 
p\left(x\right)=Cx^{-\alpha } 
\end{equation} 
where $\alpha $ is a constant parameter of distribution called the exponent or scaling parameter and \textit{C }is constant of normalization [43]. In general few quantities obeys a power law for all values of \textit{x}. Typically there is a parameter $x_{min}$ above which the values of \textit{x} follows a power law [44].

\noindent 

\noindent Power law distributions are of two distinct kind. Continuous distribution where the quantity of interest vary continuously and discrete distribution where the random variable takes only a discrete set of values, generally positive integers. A continuous power law distribution for \textit{x }is given by
\begin{equation} \label{GrindEQ__9_} 
p\left(x\right)dx={\mathrm{Pr} \left(x\le X<x+dx\right)\ }=\ Cx^{-\alpha }dx 
\end{equation} 
where \textit{X} is the random variable. This quantity diverges as $x\to 0$, thus implying there must be some lower bound $x_{min}$ to the power law behaviour.

\noindent 

\noindent Given $\alpha >1\ $the above equation \eqref{GrindEQ__9_} becomes
\begin{equation} \label{GrindEQ__10_} 
p\left(x\right)=\ \frac{\alpha -1}{x_{min}}(\ \frac{x}{x_{min}})^{-\alpha } 
\end{equation} 
For the discrete case, x takes a discrete set of values and the probability distribution has the form
\begin{equation} \label{GrindEQ__11_} 
p\left(x\right)={\mathrm{Pr} \left(X=x\right)\ }=\ Cx^{-\alpha } 
\end{equation} 
Similar to the continuous case there must be some $x_{min}>0$ such that the distribution does not diverges for the values of x above$\ x_{min}$. The constant of normalization can be found and the equation becomes
\begin{equation} \label{GrindEQ__12_} 
p\left(x\right)=\ \frac{x^{-\alpha }}{\zeta (\alpha ,\ x_{min})} 
\end{equation} 
where
\begin{equation} \label{GrindEQ__13_} 
\zeta \left(\alpha ,\ x_{min}\right)=\ \sum^{\infty }_{n=0}{(n+\ x_{min}})^{-\alpha } 
\end{equation} 
is the generalized zeta function.

\noindent It must be noted that for a power law distribution if we take logarithm on both sides the equation \eqref{GrindEQ__8_} becomes
\begin{equation} \label{GrindEQ__14_} 
{\mathrm{ln} p\left(x\right)=\ -\alpha {\mathrm{ln} \left(x\right)\ }+\mathrm{ln}\mathrm{}(C)\ } 
\end{equation}

\noindent Thus the most common way to establish that given empirical data has power law or not is by plotting the data on a doubly logarithmic scale and observing if the data follows a straight line or not. A network is called a scale-free network if its degree sequence follows a power law distribution. For a scale-free network the degree sequence of the underlying graph is plotted on a doubly logarithmic scale. If the values of the distribution indeed comes from a power law then the double logarithmic plot will be a straight line. The slope of the straight line gives us some estimation of the value of scaling parameter $\alpha $. Otherwise we have to use maximum likelihood estimation to give an estimate to the value of $\alpha $. For discrete case MLE is given by the solution of equations with $x_{min}>1$ as
\begin{equation} \label{GrindEQ__15_} 
\frac{\zeta '(\alpha ,\ x_{min})}{\zeta (\alpha ,\ x_{min})} = -\ \frac{1}{n}\ \sum^n_{i=1}{{lnx}_i} 
\end{equation}

\noindent In discrete probability distribution case discussed above we assumed that the value of $x_{min}$\textit{ }is given.   However when $x_{min}$\textit{ }is not given, it can be estimated empirically by choosing an estimate that makes the cumulative probability distribution $S\left(x\right)$ and the best-fit power law model as similar as possible. The Kolmogorov -- Smirnov statistic between two probability distribution is 
\begin{equation} \label{GrindEQ__16_} 
D=max_{x\ \ge \ x_{min}}|S\left(x\right)-P\left(x\right)| 
\end{equation} 
where $P\left(x\right)\ $is the CDF for the power law model that best fits the data in region $x\ge \ x_{min}$.

\vspace{0.5 cm}
\noindent 

\noindent 

\noindent 

\newpage

\noindent \textbf{3.3 Identification of small-world networks}

\noindent 

\noindent Clustering coefficient of a vertex in a network is the ratio of number of edges present among the adjacent vertices of the vertex and total number of edges possible among the adjacent vertices. Let \textit{G }= (\textit{V,E}) be a graph. The clustering coefficient of a vertex \textit{u }with set of adjacent vertices $N_u$\textit{ }is given by
\begin{equation} \label{GrindEQ__17_} 
C\left(u\right)=\ \frac{2\ |\ e_{ij}:i,\ j\in N_u,\ e_{ij}\in E\ |}{k\ (k-1)} 
\end{equation} 
Where k is the number of vertices in $N_u$.

\noindent A similar idea is that of transitivity which is the ratio of three times the number of triangles present in a graph and number of connected triplets (L-shaped formations) of vertices in the graph [45]. Transitivity is defined as
\begin{equation} \label{GrindEQ__18_} 
T=\ \frac{3\times number\ of\ triangles}{number\ of\ paths\ of\ length\ 2} 
\end{equation}

\noindent A network is said to possess small-world property if any vertex in the network can be reached by any other vertex in the network by traversing a path consisting of only a small number of vertices. It is observed that for a network \textit{G }of order \textit{n} and size \textit{m}, the average shortest path length is similar to an Erdos - R\'{e}nyi random network of same size and order. But the transitivity $T_G$\textit{ }and clustering coefficient $C_G$\textit{ }of the network is much higher than that for the Erdos - R\'{e}nyi random network $T_{ER}$\textit{ }and $C_{ER}$ [45].\textit{ }This property is used to define a measure of small-world property. We call this as \textit{small-world }$\mathrm{-}$ \textit{ness }[46] , which is defined as
\begin{equation} \label{GrindEQ__19_} 
SW=\ \frac{T_{G\ }\times \ L_{ER}}{L_G\times T_{ER}} 
\end{equation} 
where $L_G$ and $L_{ER}$ are average shortest path length for the given network \textit{G }and Erdos - R\'{e}nyi random network of same size and order. We call a network to be small-world if the value of small-world-ness is greater than one. 
\\
\noindent 

\noindent 

\noindent \textbf{3.4 Assortative mixing}

\noindent A network is said to show assortative mixing if vertices of higher degree are connected to vertices of higher degree and the vertices of lower degree tend to connect to vertices of lower degree. A network is called disassortative if the high degree vertices are connected to vertices of low degree. A measure for assortative mixing has been proposed [47]. In a network with $N\ $vertices and $M$ edges suppose we choose an edge \textit{e }and arrive at a vertex \textit{v} along the chosen edge \textit{e.} Then the distribution of the remaining degree (the number of edges leaving the vertex other than the one along which one arrived) is given by
\begin{equation} \label{GrindEQ__20_} 
q_k=\ \frac{\left(k+1\right)p_{k+1}}{{\mathrm{\Sigma }}_j\ j\ p_j} 
\end{equation} 
where $p_k\ $is the probability that a vertex chosen at random in a network has degree equal to $k$.

\noindent Let $e_{jk}$ be the joint probability distribution of the remaining degrees of the two vertices at either end of a randomly chosen edge. Then the assortative mixing (\textit{r}) is given by 
\begin{equation} \label{GrindEQ__21_} 
r=\ \frac{1}{{\sigma }^2_q}\ {\mathrm{\Sigma }}_{jk}\left(e_{jk}-\ q_jq_k\right) 
\end{equation} 
where ${\sigma }^2_q=\ {\mathrm{\Sigma }}_kk^2q_k-[\ {\mathrm{\Sigma }}_k\ k\ q_k]^2$ is the variation of $q_k.$

\noindent For the purpose of computations, assortative mixing is calculated as
\begin{equation} \label{GrindEQ__22_} 
r=\ \frac{M^{-1}\ {\mathrm{\Sigma }}_i{\ j}_i\ k_i-[\ M^{-1}\ {\mathrm{\Sigma }}_i\ \frac{1}{2}\ (j_i+\ k_i)]^2}{\ M^{-1}\ {\mathrm{\Sigma }}_i\ \frac{1}{2}\ \left(j^2_i+\ k^2_i\right)-\ \ [M^{-1}\ {\mathrm{\Sigma }}_i\ \frac{1}{2}\ (j_i+\ k_i)]^2} 
\end{equation} 
where $j_{i\ }$and $k_i$ are the degrees of the vertices at the end of the \textit{i}th edge where $i=1\dots M$.
\\
\noindent 

\newpage

\noindent \textbf{\textsc{3. Results and Discussion}}

\noindent This distribution of the natural forest patches in the western Himalaya of India is given in figure 1.

\noindent \includegraphics*[scale=1]{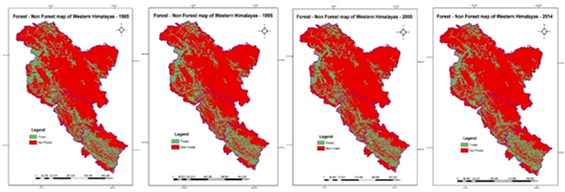}

\noindent \textit{Figure 1: Potential distribution of forest patches in the Western Himalaya region}

\vspace{0.5 cm}

\noindent We extracted and studied networks for the seed dispersion network of the mentioned anemochory species of Western Himalayas. The networks for the observed years 1985, 1995 and 2005 vary in size and order where the number of vertices represent the total number of patches present during the given period. The underlying graph of seed dispersal network for the year 1985 is formed of 2819 vertices and 3304 edges. The number of connected components are found to be equal to 546 of which 413 are isolated vertices with no edges. The giant component consists of 1600 vertices while the second largest component is of size 256 vertices. 

\noindent For the year 1995 the network consists of 2909 vertices and 3386 edges and a total of 559 connected components out of which 420 are isolated vertices or habitat patches which are not close enough to other habitat patches such that ecological information (seeds) can flow from one patch to another. Again a threshold of three hundred meters is used to determine if ecological information can pass between habitat patches or not. The largest component is a giant component containing 1920 vertices.

\noindent 

\noindent The network for the year 2005 contains a total of 3576 vertices and 4678 edges. The network has a total of 610 components out of which 447 vertices are isolated vertices i.e. components of size one. Thus a total of 163 components contains at least one edge. The biggest component is a giant component of size equal to 1905 vertices and the second largest component consists of 404 vertices. Communities in a network are group of vertices that are present such that they have more edges linking the vertices of the group as compared to vertices outside of the group. Which means that the group forms a local cluster and has a locally cliquish topology. We used the Newman - Girvan algorithm to find communities in network and found that a total of 189 communities are present in the network. This means that some components contain more than one communities [48]. In terms of seed dispersal this could indicate that there exist patches which are closely knit such that seeds from several neighbouring patches accumulate in every patch of the community because of the locally cliquish nature of communities. In a given large component there are more than one location where such a phenomenon occurs. 
\\
\noindent 

\newpage

\noindent We further computed centralities for the network. For the year 1985 the different centralities rank the vertices as follows.

\noindent 

\begin{tabular}{|p{0.5in}|p{0.4in}|p{0.5in}|p{0.4in}|p{0.4in}|p{0.5in}|p{0.4in}|p{0.5in}|p{0.4in}|p{0.5in}|} \hline 
\multicolumn{2}{|p{1in}|}{Degree Centrality  ${(C}_D)$} & \multicolumn{2}{|p{0.9in}|}{Eigenvector Centrality  ${(C}_E)$} & \multicolumn{2}{|p{0.9in}|}{Betweenness Centrality  ${(C}_B)$} & \multicolumn{2}{|p{0.9in}|}{Subgraph Centrality  ${(C}_S)$} & \multicolumn{2}{|p{0.9in}|}{Closeness Centrality ${(C}_C)$} \\ \hline 
Vertex No.  & Degree & Vertex No. & Value & Vertex No. & Value & Vertex No. & Value & Vertex No. & Value \\ \hline 
1811 & 200 & 1811 & 0.6808 & 1811 & 1.72E+06 & 1811 & 1.51E+06 & 1811 & 536.39 \\ \hline 
1572 & 106 & 2339 & 0.0803 & 484 & 1.05E+06 & 1572 & 2.54E+04 & 1572 & 460.15 \\ \hline 
1801 & 66 & 1694 & 0.0761 & 1572 & 1.05E+06 & 1694 & 2.16E+04 & 1694 & 456.53 \\ \hline 
598 & 61 & 2238 & 0.0720 & 1694 & 9.93E+05 & 2339 & 2.14E+04 & 1813 & 408.68 \\ \hline 
1309 & 61 & 2172 & 0.0686 & 1813 & 9.70E+05 & 2238 & 1.71E+04 & 1552 & 398.62 \\ \hline 
1694 & 57 & 2241 & 0.0677 & 443 & 8.98E+05 & 2172 & 1.54E+04 & 1561 & 393.85 \\ \hline 
778 & 50 & 2288 & 0.0671 & 598 & 5.42E+05 & 2241 & 1.50E+04 & 351 & 390.94 \\ \hline 
830 & 40 & 1737 & 0.0667 & 1788 & 4.64E+05 & 2288 & 1.47E+04 & 1702 & 390.94 \\ \hline 
1484 & 35 & 1735 & 0.0655 & 2356 & 3.67E+05 & 1737 & 1.47E+04 & 2339 & 388.36 \\ \hline 
484 & 34 & 1813 & 0.0654 & 1769 & 3.65E+05 & 1813 & 1.41E+04 & 1737 & 382.70 \\ \hline 
\end{tabular}

\noindent \includegraphics*[width=6.49in, height=5.46in, keepaspectratio=false]{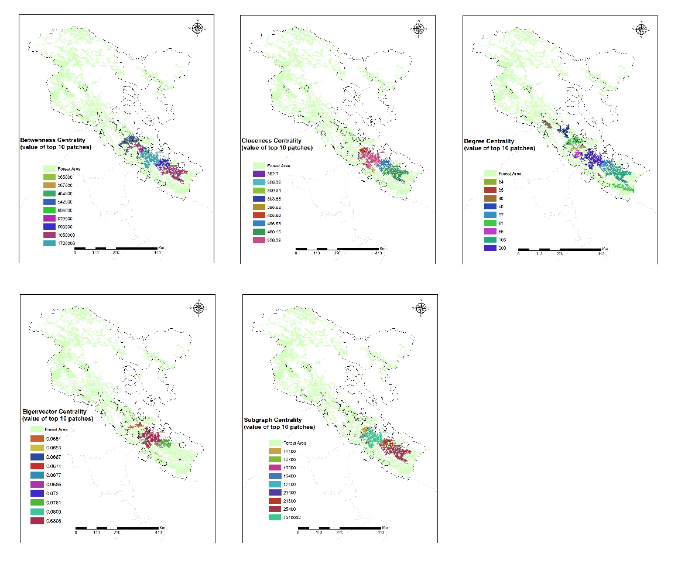}

\noindent \textit{Figure 2: Maps for various centrality indices computed for forest cover of 1985}

\vspace{0.5 cm}

\newpage

\noindent For the year 1995 the following rankings as per the different centralities are obtained.

\noindent 

\begin{tabular}{|p{0.5in}|p{0.4in}|p{0.5in}|p{0.4in}|p{0.4in}|p{0.5in}|p{0.4in}|p{0.5in}|p{0.4in}|p{0.5in}|} \hline 
\multicolumn{2}{|p{1in}|}{Degree Centrality  ${(C}_D)$} & \multicolumn{2}{|p{0.9in}|}{Eigenvector Centrality  ${(C}_E)$} & \multicolumn{2}{|p{0.9in}|}{Betweenness Centrality  ${(C}_B)$} & \multicolumn{2}{|p{0.9in}|}{Subgraph Centrality  ${(C}_S)$} & \multicolumn{2}{|p{0.9in}|}{Closeness Centrality  ${(C}_C)$} \\ \hline 
Vertex No.  & Degree & Vertex No. & Value & Vertex No. & Value & Vertex No. & Value & Vertex No. & Value \\ \hline 
1632 & 106 & 1909 & 0.530 & 1909 & 2.18E+06 & 1909 & 3.29E+04 & 1909 & 489.41 \\ \hline 
1909 & 105 & 1754 & 0.405 & 562 & 1.87E+06 & 1754 & 2.53E+04 & 1632 & 466.52 \\ \hline 
1754 & 98 & 1773 & 0.136 & 1911 & 1.67E+06 & 1632 & 2.50E+04 & 1773 & 453.96 \\ \hline 
853 & 62 & 1632 & 0.126 & 684 & 1.43E+06 & 1773 & 4.30E+03 & 1754 & 449.72 \\ \hline 
684 & 61 & 2333 & 0.117 & 1632 & 1.35E+06 & 1369 & 2.49E+03 & 1911 & 404.58 \\ \hline 
1369 & 61 & 2331 & 0.092 & 1773 & 1.33E+06 & 853 & 2.12E+03 & 1621 & 391.29 \\ \hline 
1880 & 59 & 1714 & 0.092 & 2575 & 1.30E+06 & 1880 & 2.02E+03 & 2333 & 390.55 \\ \hline 
1773 & 57 & 2407 & 0.088 & 853 & 1.10E+06 & 684 & 1.67E+03 & 424 & 383.80 \\ \hline 
929 & 40 & 1795 & 0.082 & 1754 & 1.06E+06 & 2333 & 1.45E+03 & 1782 & 383.80 \\ \hline 
1544 & 35 & 1911 & 0.075 & 894 & 8.31E+05 & 2407 & 1.18E+03 & 1612 & 383.74 \\ \hline 
\end{tabular}

\noindent \includegraphics*[width=6.48in, height=5.61in, keepaspectratio=false]{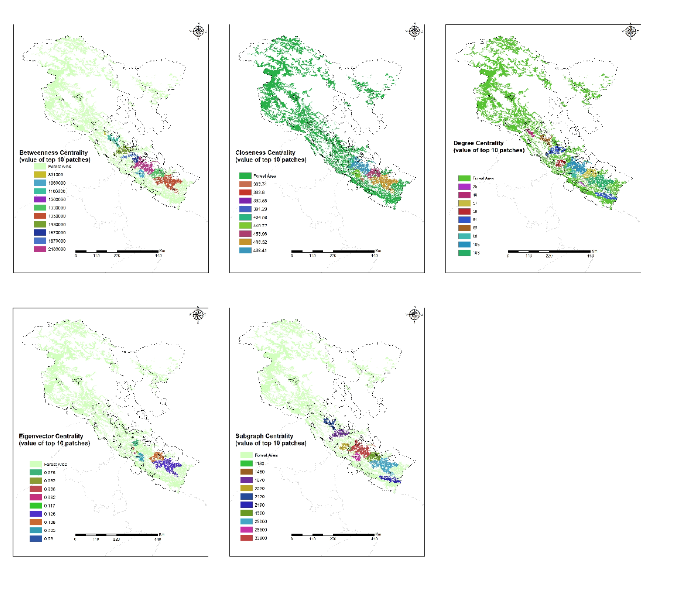}

\noindent \textit{Figure 3: Maps for various centrality indices computed for forest cover of 1995}

\vspace{0.5 cm}

\newpage

\noindent For the year 2005 the rankings and values of different centralities are found as follows.

\begin{tabular}{|p{0.5in}|p{0.4in}|p{0.5in}|p{0.4in}|p{0.4in}|p{0.5in}|p{0.4in}|p{0.5in}|p{0.4in}|p{0.5in}|} \hline 
\multicolumn{2}{|p{1in}|}{Degree Centrality  ${(C}_D)$} & \multicolumn{2}{|p{0.9in}|}{Eigenvector Centrality  ${(C}_E)$} & \multicolumn{2}{|p{0.9in}|}{Betweenness Centrality  ${(C}_B)$} & \multicolumn{2}{|p{0.9in}|}{Subgraph Centrality  ${(C}_S)$ } & \multicolumn{2}{|p{0.9in}|}{Closeness Centrality  ${(C}_C)$} \\ \hline 
Vertex No.  & Degree & Vertex No. & Value & Vertex No. & Value & Vertex No. & Value & Vertex No. & Value \\ \hline 
2218 &  106 & 2218 & 0.5705 & 2478 & 2.05E+06 & 2218 & 2.50E+04 & 2478 & 4.56E+02 \\ \hline 
2341 & 89 & 2478 & 0.2588 & 1253 & 1.93E+06 & 2341 & 1.53E+04 & 2218 & 4.55E+02 \\ \hline 
2478 & 86 & 2341 & 0.2143 & 2480 & 1.67E+06 & 2478 & 1.37E+04 & 2360 & 4.36E+02 \\ \hline 
1452 & 85 & 2360 & 0.2126 & 1122 & 1.51E+06 & 1452 & 8.77E+03 & 2341 & 4.26E+02 \\ \hline 
1955 & 61 & 2105 & 0.1126 & 1130 & 1.51E+06 & 2360 & 3.76E+03 & 2207 & 3.74E+02 \\ \hline 
2360 & 57 & 2007 & 0.0912 & 2218 & 1.33E+06 & 1955 & 2.49E+03 & 2049 & 3.74E+02 \\ \hline 
1625 & 54 & 2754 & 0.0829 & 2360 & 1.30E+06 & 1625 & 1.37E+03 & 2480 & 3.72E+02 \\ \hline 
532 & 41 & 2178 & 0.0794 & 1452 & 1.12E+06 & 2105 & 1.11E+03 & 2897 & 3.71E+02 \\ \hline 
2455 & 37 & 2191 & 0.0787 & 1241 & 9.99E+05 & 532 & 1.03E+03 & 2198 & 3.70E+02 \\ \hline 
2130 & 35 & 2630 & 0.0732 & 1030 & 9.73E+05 & 2897 & 899.275 & 2754 & 367.4525 \\ \hline 
\end{tabular}

\noindent \includegraphics*[width=6.53in, height=5.28in, keepaspectratio=false, trim=0.00in 0.39in 0.00in 0.00in]{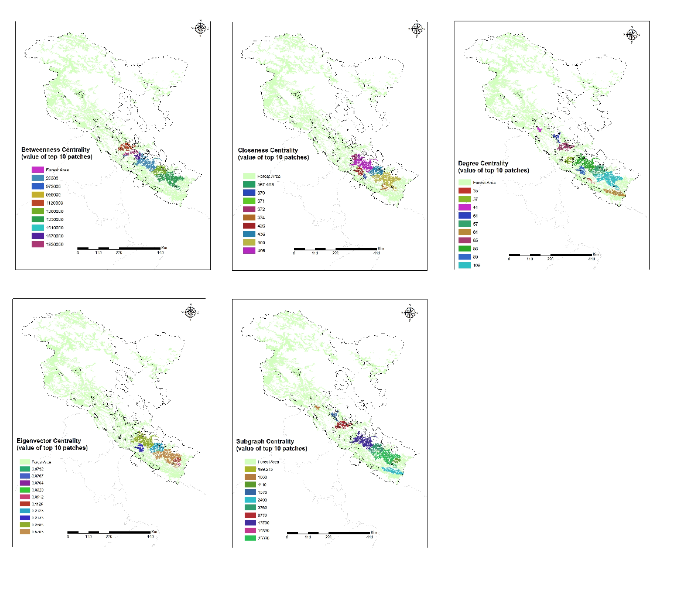}

\noindent \textit{Figure 4: Maps for various centrality indices computed for forest cover of 2005}

\vspace{0.5 cm}

\noindent The forest patches that are best ranked according to different centrality measures are important for the purpose of conservation as they demand special consideration for the role that they play in the process of dispersal of the anemochory plant species \textit{Abied pindrow, Betula utilis} and \textit{Taxus wallichiana} in the Western Himalaya region.

\noindent The Pearson correlation coefficient for two random variables \textit{X }and \textit{Y }is given by the covariance of the two variables divided by the product of their standard deviation. For the set of observations $X=\left\{x_1,x_2,\ \dots ,\ x_n\right\}\ $and $Y=\left\{y_1,y_2,\ \dots ,\ y_n\right\}$ the Pearson correlation coefficient is given by
\begin{equation} \label{GrindEQ__23_} 
\rho =\ \frac{{\mathrm{\Sigma }}_i(x_i-\ \overline{x})(y_i-\ \overline{y})}{\sqrt{{\mathrm{\Sigma }}_i\ (x_i-\ \overline{x})^2}\ \ \sqrt{{\mathrm{\Sigma }}_i(y_i-\ \overline{y})^2}} 
\end{equation} 
\\
The correlation between ten top ranked vertices has been calculated for the network of year 2005 and the following values are obtained.

\noindent 

\begin{tabular}{|p{0.7in}|p{0.7in}|p{0.7in}|p{0.7in}|p{0.7in}|p{0.7in}|} \hline 
 & $C_D$ & $C_E$ & $C_B$ & $C_S$ & $C_C$ \\ \hline 
$C_D$ & 1 &  &  &  &  \\ \hline 
$C_E$ & -0.09794 & 1 &  &  &  \\ \hline 
$C_B$ & 0.271083 & 0.178529 & 1 &  &  \\ \hline 
$C_S$ & -0.02592 & -0.05227 & 0.304326 & 1 &  \\ \hline 
$C_C$ & -0.81289 & 0.207835 & 0.093546 & 0.160646 & 1 \\ \hline 
\end{tabular}

\vspace{0.5 cm}

\noindent 

\noindent It is observed that generally there is a lack of correlation between top ranked vertices as calculated for various centrality measures. However the closeness centrality and degree centrality are strongly anti-correlated. For the network of year 1995 the correlation between top ranked vertices is given as

\noindent 

\noindent 

\begin{tabular}{|p{0.7in}|p{0.7in}|p{0.7in}|p{0.7in}|p{0.7in}|p{0.7in}|} \hline 
 & $C_D$ & $C_E$ & $C_B$ & $C_S$ & $C_C$ \\ \hline 
$C_D$ & 1 &  &  &  &  \\ \hline 
$C_E$ & -0.13654 & 1 &  &  &  \\ \hline 
$C_B$ & 0.06975 & -0.03128 & 1 &  &  \\ \hline 
$C_S$ & -0.08749 & -0.75943 & 0.07083 & 1 &  \\ \hline 
$C_C$ & -0.17746 & -0.56219 & 0.58699 & 0.57862 & 1 \\ \hline 
\end{tabular}

\vspace{0.5 cm}

\noindent 

\noindent It is found that there is good correlation between betweenness centrality and closeness centrality and closeness centrality and subgraph centrality. The eigenvector centrality and subgraph centrality and the closeness centrality and eigenvector centrality show a strong negative correlation. The correlation between top ranked vertices has been calculated for the network of the year 1985. The following values are found.

\noindent 

\noindent 

\begin{tabular}{|p{0.7in}|p{0.7in}|p{0.7in}|p{0.7in}|p{0.7in}|p{0.7in}|} \hline 
 & $C_D$ & $C_E$ & $C_B$ & $C_S$ & $C_C$ \\ \hline 
$C_D$ & 1 &  &  &  &  \\ \hline 
$C_E$ & -0.11828 & 1 &  &  &  \\ \hline 
$C_B$ & -0.12897 & -0.73541 & 1 &  &  \\ \hline 
$C_S$ & -0.51854 & 0.35215 & -0.05926 & 1 &  \\ \hline 
$C_C$ & 0.26775 & -0.54427 & 0.66588 & -0.37857 & 1 \\ \hline 
\end{tabular}
\\
\vspace{0.5 cm}

\noindent There is a strong correlation between closeness centrality and betweenness centrality where as a strong negative correlation is found between eigenvector centrality and betweenness centrality, degree centrality and subgraph centrality and eigenvector centrality and closeness centrality.

\noindent 

\noindent We calculated assortativity for seed dispersal network and found the value to be -0.1146, -0.1417 and -0.1073 for the networks of the year 2005, 1995 and 1985 respectively. A negative value of assortitative mixing across the networks indicate that in seed dispersal network the higher degree vertices are mostly connected to vertices of low degree. Thus we infer the existence of patches that lie within the threshold distance of several other patches. However, we also infer that patches that are close to several other patches are not close to each other.

\noindent 

\noindent We use maximum likelihood estimation as given by equation \eqref{GrindEQ__15_} to estimate the value of exponent $\alpha $ for the given network and found the estimated values to be equal to 3.13, 2.92 and 2.88 respectively for the networks of year 2005, 1995 and 1985.  Also we found $x_{min}$\textit{ }= 4 for network of year 2005 and $x_{min}=3\ $for other two networks as estimated by the method of Kolmogorov - Smirnov statistic given by equation \eqref{GrindEQ__16_}. 

\noindent The values of $x>\ x_{min}$\textit{ }and corresponding degree sequence of the underlying graphs for the given network was plotted on a doubly logarithmic scale. The following plots were obtained for the subsequent years. 

\noindent 

\noindent \includegraphics*[scale=0.62]{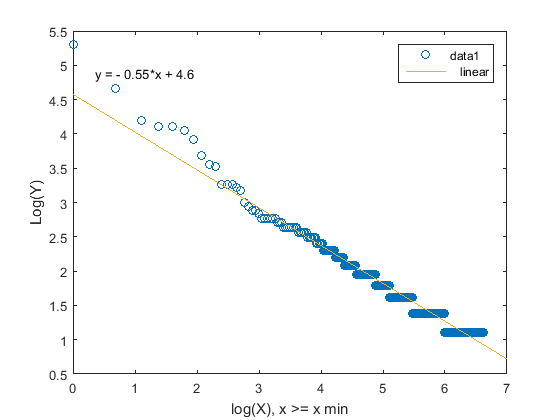}

\noindent \textit{Figure 5 Scale-free network for the year 1985}

\noindent 

\noindent \includegraphics*[scale=0.62]{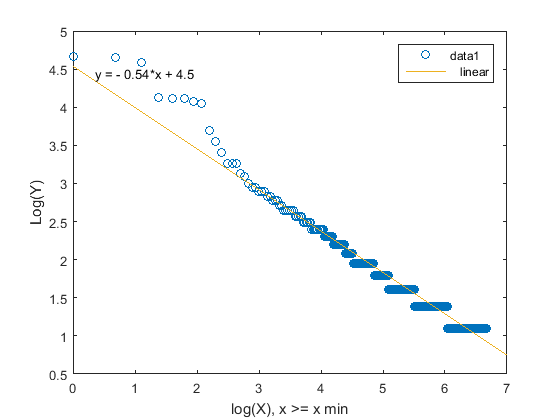}

\noindent \textit{Figure 6 Scale-free network for the year 1995}

\noindent 

\noindent \includegraphics*[scale=0.48]{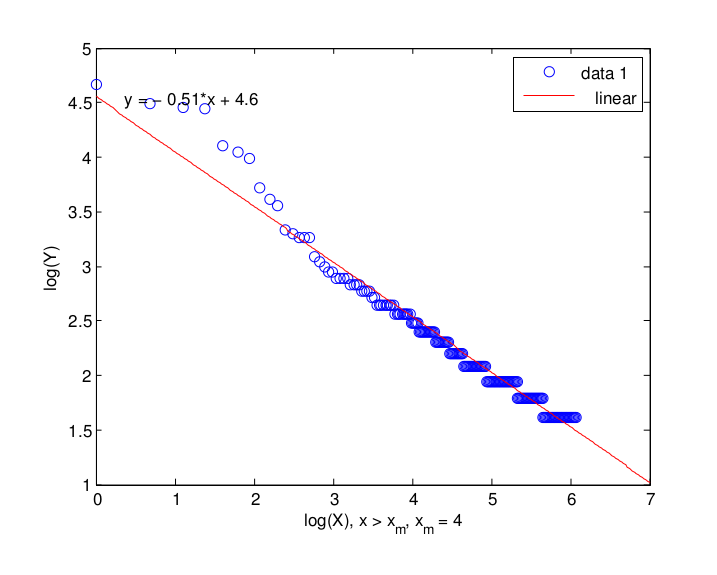}

\noindent \textit{Figure 7 Scale-free network for the year 2005}

\vspace{1 cm}

\noindent The degree sequence as plotted on a doubly logarithmic scale for values of $x>\ x_{min}$ follows a straight line in each case thus confirming that the seed dispersal network of given species of the Western Himalayas is a scale free network. Hubs in a scale-free network are the vertices with largest degree. For the network corresponding to year 2005, we calculated the degree share of hubs which is the ratio of degree of a vertex to total number of vertices in the network and found that the vertex with highest degree has 2.96 \% of degree share. Moreover the top ten highest degree vertices has an aggregate 18.2 \% of degree share. Thus in the network a large number of vertices are incident to only a few vertices with high degree which is typical of a scale-free network. For the network of habitat patches this would mean that there is a very less possibility of the random extinction of plant species in a hub patch as there are many patches with a few patches or no patches close to them. This would also mean that once the plant species thrives in a hub patch it could further spread from there to patches adjacent to hub patch by dispersal and thus thrive in a large region.

\noindent 

\noindent Furthermore, for the network for year 2005 we found the value of average clustering coefficient to be 0.3220 and transitivity to be equal to 0.1291. We found the value of clustering coefficient and transitivity for the Erdos - R\'{e}nyi random network of same size and order to be 0.0010 and 0.0017 respectively which is far less than that observed for the given network. The average shortest path length for given network is 3.1500 and same for the Erdos - R\'{e}nyi random network is 6.7248. Thus the value of small-world-ness for given network is 161.4958 which shows that the given network has the small-world property. High clustering in a scale-free networks indicate the presence of small dense subgraphs in which the adjacent vertices of a vertex are themselves adjacent thus forming locally cliquish subgraphs of small size ad order. Such small subgraphs in a scale-free network are connected together by a hub which is a vertex of high degree.

\noindent 

\noindent 

\noindent 

\noindent A scale-free network with high clustering is said to show hierarchical organization and hence modularity if the clustering coefficient varies as the inverse of degree for a vertex as $C_G\propto \ \frac{1}{d}$\textit{ }, where \textit{d }is the degree of vertex \textit{x}. [40] For the dispersal network corresponding to year 2005 we found no such scaling and the value of linear correlation coefficient between clustering coefficient and degree to be equal to 0.0778 thus indicating the absence of any linear correlation between the two quantities. This could be because the seed dispersal network shows geographical organization as this is one primary reason why hierarchical organization is not found in a scale-free network with high clustering as established by Barabasi and Ravasz. [49] 

\noindent 

\noindent The value of average clustering coefficient for network corresponding to year 1995 is equal to 0.2698 while the transitivity of the network is 0.0784. The average shortest distance for the same network is found to be 4.9153 while the transitivity for Erdos - R\'{e}nyi random network of same size and order is 3.74E-04 whereas the average shortest distance for Erdos - R\'{e}nyi random network is 6.5859. Thus the network shows small-world property as the value of small-world-ness as given by equation \eqref{GrindEQ__19_} is found to be equal to 280.7912.

\noindent 

\noindent For the network for the year 1985 the value of average clustering coefficient and transitivity are found to be equal to 0.2766 and 0.0637. The average shortest distance for the network is 2.7654. The transitivity and average shortest distance for Erdos - R\'{e}nyi random network of same size and order is found to be equal to 7.64E-04 and 6.7265. Thus on comparison to Erdos - R\'{e}nyi random network the network is found to possess the small-world property with the value of small-world-ness to be equal to 202.6398. 

\noindent 

\noindent The network being small-world implies that perhaps even being sparsely connected, there is high clustering at the local scale together with short average path length. For the process of seed dispersal this would mean that seeds from a forest patch in which the anemochory plant species are evenly present will disperse to neighbouring patches and since the neighbouring patches are themselves neighbours (presence of triangles in the network) the propagation of gene pool will take place evenly on a local scale. Because of the small-world property, an interesting network dynamics unfolds in the scenario when the species are introduced in a connected component of the network. Due to clustering the species when introduced to a forest patch will showcase an even spread locally in the forest component. This is because locally there is mutual dispersion from among the members of a cluster. Also, due to low average shortest path length in the network the spread of plant species becomes possible to relatively distant patches in the component as all the patches are only some steps away from each other. Thus due to mesoscale transitions the spread of species happens rapidly in a component on an ecological time scale. Moreover if a hub is present in the component then there is a sudden burst in the spread of the species when the species conquers a hub patch and begin dispersion from there. Thus the species when introduced to a connected component is likely to thrive because of scale-free and small-world nature of the network. 

\noindent 
\\
\noindent \textbf{\textsc{References}}
\\
\noindent 

\noindent [1] Bascompte, J., Jordano, P., Melian, C. J. \& Olesen, J. M. The nested assembly of plant–animal mutualistic networks. \textit{Proc. Natl. Acad. Sci. U. S. A.} 100, 9383--9387, 2003

\noindent [2] Olesen, J. M., Stefanescu, C. \& Traveset, A. Strong, long-term temporal dynamics of an ecological network. \textit{PLoS One 6}, e26455, 2011.

\noindent

\noindent [3] Jordano, P., Bascompte, J. \& Olesen, J. M. Invariant properties in coevolutionary networks of plant-animal interactions. \textit{Ecol. Lett. 6}, 69--81, 2003.

\noindent

\noindent [4] Thebault, E. \& Fontaine, C. Stability of ecological communities and the architecture of mutualistic and trophic networks. \textit{Science} 329, 853--856, 2010.

\noindent

\noindent [5] Bascompte, J. \& Jordano, P. Plant-animal mutualistic networks: The architecture of biodiversity. \textit{Annu. Rev. Ecol. Evol. Syst.} 38, 567--593, 2007.

\noindent

\noindent [6] Ings, T. C. et al. Ecological networks–beyond food webs. \textit{J. Anim. Ecol.} 78, 253--269, 2009.

\noindent

\noindent [7] Olesen, J. M., Bascompte, J., Dupont, Y. L. \& Jordano, P. The modularity of pollination networks. \textit{Proc. Natl. Acad. Sci. U. S. A.} 104, 19891--19896, 2007.

\noindent

\noindent [8] Bascompte, J. Disentangling the web of life. \textit{Science} 325, 416--419, 2009.

\noindent

\noindent [9] Pilosof S, Greenbaum G, Krasnov BR, Zelnik Y. Asymmetric disease dynamics in multihost interconnected networks \textit{arXiv:1512.09178} 2016.

\noindent [10] Cantwell, M.D., Forman, R.T. Landscape graphs: ecological modeling with graph theory to detect configurations common to diverse landscapes. \textit{Landsc. Ecol.} 8 \eqref{GrindEQ__4_}, 239--255, 1993.

\noindent 

\noindent [11] Fall, A., Fortin, M.J., Manseau, M., O'Brien, D. Spatial graphs: principles and applications for habitat connectivity. \textit{Ecosystems} 10 \eqref{GrindEQ__3_}, 448--461, 2007.

\noindent 

\noindent [12] Gastner, M.T., Newman, M.E.J. Optimal design of spatial distribution networks. \textit{Phys. Rev. E: Stat. Nonlinear Soft Matter Phys.} 74 \eqref{GrindEQ__1_}, 016117. 2006.

\noindent 

\noindent [13] Minor, E.S., Urban, D.L. Graph theory as a proxy for spatially explicit population models in conservation planning. \textit{Ecol. Appl.} 17 \eqref{GrindEQ__6_}, 1771--1782, 2007.

\noindent 

\noindent [14] Minor, E.S., Urban, D.L. A graph-theory framework for evaluating landscape connectivity and conservation planning. \textit{Conserv. Biol.} 22 \eqref{GrindEQ__2_}, 297--307, 2008.

\noindent 

\noindent [15] Urban, D., Keitt, T. Landscape connectivity: a graph-theoretic perspective. \textit{Ecology} 82 \eqref{GrindEQ__5_}, 1205--1218, 2001.

\noindent 

\noindent [16] Chetkiewicz, C.L.B., St Clair, C.C., Boyce, M.S. Corridors for conservation: integrating pattern and process. \textit{Annu. Rev. Ecol. Evol. Syst.} 37, 317--342, 2006.

\noindent [17] Taylor, P. D., Fahrig, L., With, K. A. Landscape connectivity: a return to the basics. In: Connectivity conservation, Crooks, K.R. \&amp; Sanjayan, M. (eds.), pp.29 - 43, Cambridge University Press, Cambridge. 2006.

\noindent 

\noindent 

\noindent [18] Crooks, K. R., \& Sanjayan, M. Connectivity conservation: maintaining connections for nature. \textit{Conservation Biology Series}-Cambridge, \textit{14}, 1. 2006.

\noindent [19] Patten, B.C. Systems approach to the concept of environment. \textit{Ohio J. Sci.} 78,206--222, 1978.

\noindent 

\noindent [20] Patten, B.C. Environs: the superniches of ecosystems. \textit{Am. Zool.} 21,845--852, 1981.

\noindent 

\noindent [21] Patten, B.C. Environs: relativistic elementary particles for ecology. \textit{Am. Nat.} 119, 179--219, 1982.

\noindent 

\noindent [22] Patten, B.C. Energy cycling in the ecosystem. \textit{Ecol. Model.} 28, 1--71, 1985.

\noindent 

\noindent [23] Fath, B.D., Patten, B.C. Review of the foundations of network environ analysis. \textit{Ecosystems} 2, 167--179, 1999.

\noindent 

\noindent [24] Roy, A., Bhattacharya, S., Ramprakash, M., Kumar, A. S. Modelling critical patches of connectivity for invasive Maling bamboo (Yushania maling) in Darjeeling Himalayas using graph theoretic approach. \textit{Ecological Modelling} 329 77--85, 2016.

\noindent 

\noindent [25] Ulanowicz, R.E. Identifying the structure of cycling in ecosystems. \textit{Math.Biosci.} 65, 219--237, 1983.

\noindent 

\noindent [26] Ulanowicz, R.E. Growth and Development: Ecosystem Phenomenology. \textit{Springer-Verlag, New York, NY} 1986.

\noindent 

\noindent [27] Ulanowicz, R.E. Ecology, the Ascendant Perspective. \textit{Columbia University Press, New York, NY} 1997.

\noindent 

\noindent [28] Ulanowicz, R.E. Quantitive methods for ecological network analysis. \textit{Comput.Biol. Chem.} 28, 321--339, 2004.

\noindent 
\noindent [29] Stohlgren, T.J., Ma, P., Kumar, S., Rocca, M., Morisette, J.T., Jarnevich, C.S., Benson,N. Ensemble habitat mapping of invasive plant species. \textit{Risk Anal.} 30 \eqref{GrindEQ__2_},224--235, 2010.

\noindent [30] Taylor, P. D., Fahrig, L., Henein, K., \& Merriam, G. Connectivity is a vital element of landscape structure. \textit{Oikos}, 571-573, 1993.

\noindent 

\noindent [31] Hanski, I., \& Gilpin, M. E. Metapopulation Biology: Ecology, Genetics, and Evolution. \textit{Academic Press, San Diego.} 1997.

\noindent 

\noindent 

\noindent [32] Vos, C. C., Baveco, H., \& Grashof-Bokdam, C. J. Corridors and species dispersal. In applying landscape ecology in biological conservation \textit{Springer New York.} 84-104, 2002.

\noindent  

\noindent [33] Vos, C. C., Verboom, J., Opdam, P. F., \& Ter Braak, C. J. Toward ecologically scaled landscape indices. \textit{The American Naturalist}, \textit{157}\eqref{GrindEQ__1_}, 24-41, 2001.

\noindent 

\noindent 

\noindent [34] Keyghobadi, N. The genetic implications of habitat fragmentation for animals. \textit{Canadian Journal of Zoology}, \textit{85}\eqref{GrindEQ__10_}, 1049-1064, 2007. 

\noindent 
\noindent [35] Bodin, \"{O}., Norberg, J. A network approach for analyzing spatially structured populations in fragmented landscape. \textit{Landsc. Ecol.} 22 \eqref{GrindEQ__1_}, 31--44, 2007.

\noindent 

\noindent [36] Estrada, E., Bodin, \"{O}. Using network centrality measures to manage landscape connectivity. \textit{Ecol. Appl. 18} \eqref{GrindEQ__7_}, 1810--1825, 2008.

\noindent [37] Hajra, P. K., \& Rao, R. R. Distribution of vegetation types in northwest Himalaya with brief remarks on phytogeography and floral resource conservation. \textit{Proceedings: Plant Sciences}, \textit{100}\eqref{GrindEQ__4_}, 263-277, 1990.

\noindent 

\noindent 

\noindent [38] Brandes, U., Erlebach, T. Network Analysis: Methodological Foundations (Lecture Notes in Computer Science). \textit{Springer-Verlag New York, Inc., Secaucus, NJ, USA} 2005.

\noindent 

\noindent [39] Newman, M. E. J. The structure and function of complex networks. \textit{SIAM Review}, 45\eqref{GrindEQ__2_}:167--256, 2003.

\noindent 

\noindent [40] Freeman, L. C. A set of measures of centrality based on betweenness. \textit{Sociometry}, 40\eqref{GrindEQ__1_}:35--41, 1977.

\noindent 

\noindent [41] Bonacich, P. Power and centrality: A family of measures. \textit{American Journal of Sociology}, 92\eqref{GrindEQ__5_}:1170--1182, 1987.

\noindent 

\noindent [42] Estrada, E. and Rodr\'{i}guez-Vel\'{a}zquez, J. A. Subgraph centrality in complex networks. \textit{Phys. Rev. E}, 71:056103, 2005.

\noindent 

\noindent [43] Clauset, A., Shalizi, C. R. and Newman, M. E. J. Power-law distributions in empirical data. \textit{SIAM Review}, 51\eqref{GrindEQ__4_}:661--703, 2009.

\noindent 

\noindent [44] Barab\'{a}si, A. and Albert, R. Emergence of scaling in random networks. \textit{Science}, 286(5439): 509--512, 1999.

\noindent 

\noindent [45] Watts, D. J. and Strogatz, S. H. Collective dynamics of 'small-world' networks. \textit{Nature}, 393: 440--442, 1998.

\noindent 

\noindent [46] Humphries, M. D. and Gurney, K. Network small-world-ness: a quantitative method for determining canonical network equivalence. \textit{PloS one}, 3\eqref{GrindEQ__4_}:e0002051, 2008.

\noindent 

\noindent 

\noindent [47] Newman, M. E. J. Assortative mixing in networks. \textit{Physical Review Letters}, 89, 2002.

\noindent [48] Newman, M. E. J. and Girvan, M. Finding and evaluating community structure in networks. \textit{Phys. Rev. E}, 69, 2003.

\noindent 

\noindent [49] Ravasz, E. and Barab\'{a}si, A. Hierarchical organization in complex networks. \textit{Phys. Rev. E}, 67:026112, 2003.

\noindent 

\noindent

\end{document}